\documentclass[aps,amsmath,amssymb,twocolumn]{revtex4}
\usepackage[utf8]{inputenc}
\usepackage{CJK}

\usepackage{graphicx}
\usepackage{bm}
\usepackage{xcolor}
\usepackage{array}
\usepackage{braket}
\providecommand{\mel}[3]{\left\langle #1 \middle| #2 \middle| #3 \right\rangle}
\usepackage{mathtools}
\usepackage{float}
\usepackage{microtype}
\usepackage{wasysym}
\renewcommand{\baselinestretch}{1.0}
\begin{document}

\renewcommand{\baselinestretch}{1.0}
\title{Probing Operator Spreading with Local Quantum Fluctuations}

\author{Bert\'{u}lio de Lima Bernardo}%
 \email{bertulio.fisica@gmail.com}
\affiliation{%
Departamento de F\'{i}sica, Universidade Federal da Para\'{i}ba, 58051-900 Jo\~{a}o Pessoa, PB, Brazil
}%

\date{\today}

\begin{abstract}
Measuring out-of-time-ordered correlators (OTOCs)
is experimentally challenging, often requiring
effective reversal of many-body dynamics.
Here, we propose an alternative approach to probing
operator spreading through local quantum fluctuations,
using only forward evolution and local measurements.
Motivated by the Heisenberg uncertainty relation,
we introduce a differential local variance and
establish a rigorous bound connecting it to
operator noncommutativity.
For a ferromagnetic Heisenberg chain in the
single-magnon regime, we derive an exact relation
between the local variance and a squared-commutator
OTOC. In integrable and nonintegrable Ising chains,
the differential variance and OTOCs reveal
qualitatively consistent propagation fronts.
Our results establish local quantum fluctuations
as an experimentally accessible, state-dependent,
and complementary diagnostic of operator spreading.
\end{abstract}

\maketitle


\section{Introduction}

The spreading of initially localized quantum information is a central
feature of nonequilibrium many-body dynamics and is commonly described
in terms of operator growth and quantum scrambling
\cite{Cheneau2012,Landsman2019,LewisSwan2019}. Understanding this
process is important for characterizing how local information becomes
distributed across many-body degrees of freedom and for elucidating
the emergence of thermalization and quantum correlations in isolated
quantum systems \cite{Polkovnikov2011,Hosur2016}. Out-of-time-order correlators (OTOCs) have emerged as
a powerful diagnostic of this process, as they probe the growth of
noncommutativity between operators that are initially spatially
separated
\cite{Shenker2014,KitaevSuh2018,Li2017,Garttner2017}. Their behavior
therefore provides direct information about the propagation of operators and the emergence of scrambling in interacting quantum
systems \cite{Swingle2018}.

Despite their conceptual importance, OTOCs remain challenging to measure experimentally. Conventional protocols typically require an effective reversal of the many-body dynamics, together with controlled local operations, making them particularly sensitive to experimental imperfections
\cite{Li2017,Garttner2017,Swingle2018,Braum2022,Qi2026,LiYC2026}.
Control errors, imperfect reversal, and decoherence can contaminate
the measured signal and make it difficult to distinguish intrinsic
scrambling from experimental artifacts
\cite{YoshidaYao2019,Harris2022,LiYC2026}.
These challenges have motivated the development of alternative
approaches that avoid explicit time reversal or characterize
information spreading and scrambling through more accessible
observables and measurement protocols
\cite{Vermersch2019,Garcia2021,Blocher2024,Wysocki2025,Plodzien2026}.
This raises the question of whether the propagation of
an initially localized perturbation can be identified through
a simpler, directly measurable quantity.

Here, we show that local quantum fluctuations provide
an experimentally accessible signature of operator
spreading. We introduce a differential local variance
that quantifies the change in the fluctuations of a
local observable $B_j$ induced by a perturbation
applied at a distant site $i$, relative to the
unperturbed evolution. Motivated by the Heisenberg
uncertainty relation, we establish a rigorous bound
relating this quantity to the noncommutativity of
initially separated local operators.
Connections between quantum fluctuations, uncertainty
relations, and scrambling have also been explored
in stochastic operator dynamics
\cite{MartinezAzcona2023}.
In contrast, our approach relates the response of local fluctuations to operator noncommutativity using only forward time evolution and local measurements.

We demonstrate the approach in Heisenberg and Ising
spin chains. For a ferromagnetic Heisenberg chain in
the one-excitation regime, we derive an exact analytical
relation between the local variance and a
squared-commutator OTOC, showing that they identify
the same leading propagation front. We then investigate
an Ising chain in both integrable and nonintegrable
regimes, where the differential variance and OTOCs
exhibit qualitatively consistent propagation
timescales. Our results thus
establish local quantum fluctuations as a
state-dependent and experimentally accessible probe
of operator spreading.

\section{Operator Spreading and Local Uncertainty}

In an interacting quantum many-body system, a local operator generally develops support over increasingly large spatial regions under time evolution \cite{Swingle2018, Nahum2018,Keyserlingk2018}. In the Heisenberg picture,
\begin{equation}
A(t)=e^{iHt}Ae^{-iHt}
=\sum_{n=0}^{\infty}\frac{(it)^n}{n!}\mathcal{L}^n(A),
\label{eq:heisenberg_expansion}
\end{equation}
where $\mathcal{L}(\cdot)=[H,\cdot]$. For a local short-range
Hamiltonian, successive nested commutators can progressively enlarge the support of $A(t)$, giving rise to an effective light-cone structure constrained by the Lieb--Robinson bound \cite{LiebRobinson1972,Bravyi2006}.

Operator spreading can therefore be characterized through the
noncommutativity that develops between $A(t)$ and a local observable
$B$ supported at a distant site. Although $A$ and $B$ initially
commute because they have disjoint support, $[A(t),B]$ can become
nonzero as $A(t)$ develops support in the region of $B$.
Expanding the commutator gives
\begin{equation}
[A(t),B]
=
\sum_{n=0}^{\infty}
\frac{(it)^n}{n!}
[\mathcal{L}^n(A),B].
\label{eq:commutator_expansion}
\end{equation}
For finite-range interactions, the lowest-order terms vanish when $A$ and $B$ are sufficiently
far apart, explaining why noncommutativity develops later at larger distances.

A standard diagnostic of operator spreading is the squared
commutator,
\begin{equation}
C(t)=
\left\langle
[A(t),B]^\dagger[A(t),B]
\right\rangle ,
\label{eq:squared_commutator_general}
\end{equation}
which, for Hermitian unitary operators $A$ and $B$, is
related to the out-of-time-ordered correlator
$F(t)=\langle A(t)BA(t)B\rangle$ through
$C(t)=2[1-\operatorname{Re}F(t)]$
\cite{Swingle2018}.
For a pure state $|\psi\rangle$,
$C(t)=\|[A(t),B]|\psi\rangle\|^2$,
quantifying the action of the commutator on the chosen
state. At infinite temperature,
$C_{\infty}(t)=D^{-1}\operatorname{Tr}
\{[A(t),B]^\dagger[A(t),B]\}$,
where $D$ is the Hilbert space dimension, giving the
squared Hilbert--Schmidt norm of the commutator
normalized by $D$.

The expectation value of the commutator,
$\chi(t)=-i\langle[A(t),B]\rangle$,
is closely related to linear-response theory
\cite{Kubo1957}. However, unlike the squared
commutator, it can vanish even when
$[A(t),B]\neq 0$ due to cancellations arising
from symmetries or the initial state.
For a stationary reference state satisfying
$[\rho,H]=0$, these cancellations can be
understood from the spectral representation
of $\chi(t)$ \cite{SupplementalMaterial}.
Nevertheless, the commutator expectation value
connects operator noncommutativity to quantum
fluctuations through the Heisenberg uncertainty
relation,
$\Delta A(t)\,\Delta B
\geq
\frac{1}{2}
\left|
\langle[A(t),B]\rangle
\right|
=
\frac{1}{2}|\chi(t)|$.
This relation suggests that quantum fluctuations
may provide an alternative signature of operator
spreading.

\section{Local Variance as a Probe of Operator Spreading}

Consider two local observables $A_i$ and $B_j$, initially supported at spatially
separated sites $i$ and $j$, and define
$\chi_{ij}(t) =
-i\langle[A_i(t),B_j]\rangle$.
Since $[A_i,B_j]=0$ initially, $\chi_{ij}(0)=0$. As $A_i(t)$ spreads
through the system, however, it can develop support near site $j$ and become noncommuting with $B_j$. Squaring the Heisenberg uncertainty relation yields
\begin{equation}
(\Delta A_i)^2(t)(\Delta B_j)^2(t)
\geq
\frac{1}{4}|\chi_{ij}(t)|^2,
\label{eq:squared_HUR}
\end{equation}
showing that the noncommutativity generated by operator spreading
places a quantitative lower bound on the product of local variances.
A related bound holds for the squared commutator,
$C_{ij}(t)=
\langle[A_i(t),B_j]^\dagger[A_i(t),B_j]\rangle$.
Indeed, the Cauchy--Schwarz inequality implies
$C_{ij}(t)\geq|\chi_{ij}(t)|^2$.
Thus, both the variance product and the squared commutator are
bounded from below by the same commutator expectation, although
neither bound establishes a direct ordering between these two
quantities. 

The connection becomes particularly useful when the fluctuations of
$A_i(t)$ are taken into account explicitly. For any local observable
$A_i$ with nonzero variance, Eq.~\eqref{eq:squared_HUR} implies $(\Delta B_j)^2(t)
\geq
|\chi_{ij}(t)|^2 /
4(\Delta A_i)^2(t)$.
Since this inequality holds for every admissible $A_i$, the local
variance is bounded from below by the strongest uncertainty-relation
constraint within this family,
\begin{equation}
(\Delta B_j)^2(t)
\geq
\sup_{A_i}
\frac{|\chi_{ij}(t)|^2}{
4(\Delta A_i)^2(t)
},
\label{eq:optimized_HUR}
\end{equation}
where the supremum is taken over all local observables $A_i$. Thus, as initially local operators spread and develop noncommutativity with $B_j$, they generate a family of uncertainty-relation lower bounds on the fluctuations of the distant probe, with $(\Delta B_j)^2$ necessarily exceeding or equaling the strongest of these bounds.

Inequality~\eqref{eq:optimized_HUR} provides the central motivation for using local variance as a propagation diagnostic. The local variance $(\Delta B_j)^2$ is an experimentally accessible quantity that retains a rigorous connection, through the uncertainty relation, to the noncommutativity generated by operator spreading. We emphasize that this connection is one-sided: a growing commutator expectation can impose a lower bound on the variance, whereas a large variance may also arise from other sources and does not, by itself, imply operator spreading.

\begin{figure*}
    \centering
\includegraphics[width=0.85\textwidth]{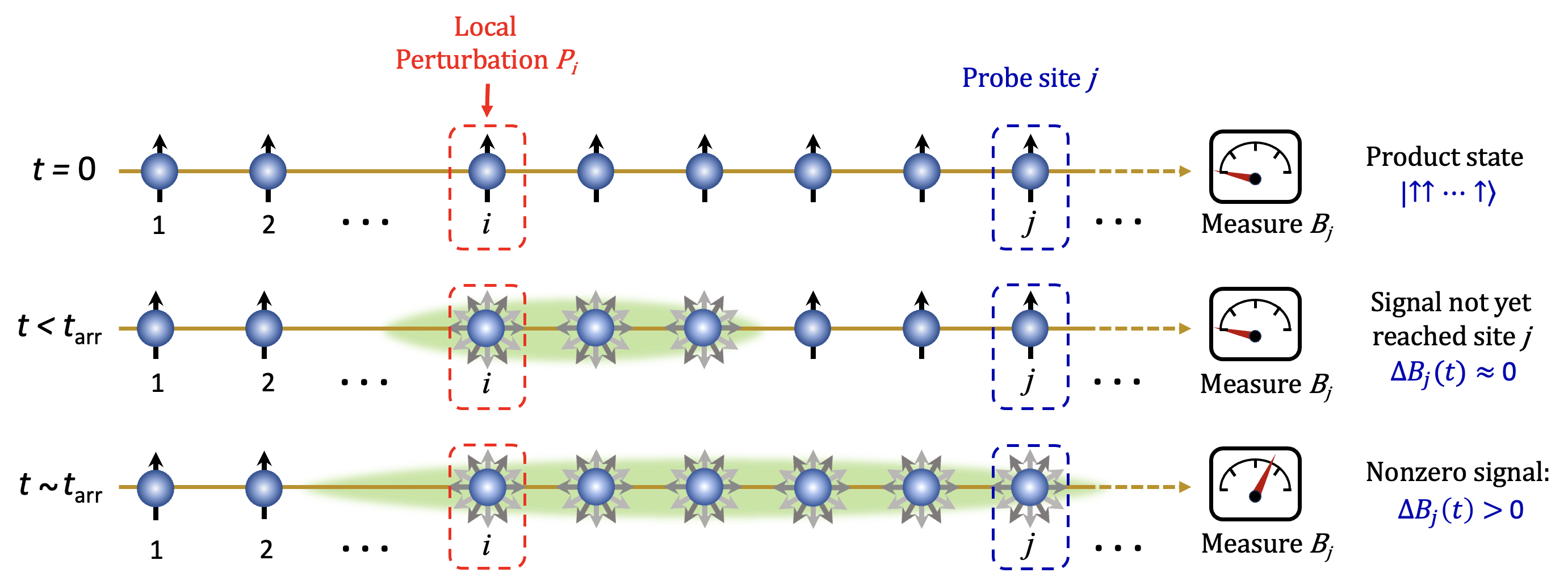}
\caption{
Protocol for probing operator spreading through local fluctuations.
Here, the system is initialized in a product state, with the probe at
site $j$ prepared in an eigenstate of the local observable $B_j$, such
that $\Delta B_j(0)=0$. A local perturbation $P_i$ is applied at a
distant site $i$ at $t=0$. Before the arrival time
$t_{\mathrm{arr}}$, the influence of the perturbation on the probe is
strongly suppressed and $\Delta B_j(t)\approx0$. As the perturbation
spreads through the system and reaches site $j$ at
$t\sim t_{\mathrm{arr}}$, the local fluctuations develop a finite
response, $\Delta B_j(t)>0$. The onset of this response provides a
local signature of the arrival of the perturbation and, consequently,
of operator spreading.
}
\label{scheme}
\end{figure*}

This observation suggests a simple protocol, illustrated schematically
in Fig.~\ref{scheme}. Prepare a state in which
the probe at site $j$ is initially an eigenstate of $B_j$, so that$
(\Delta B_j)^2(0)=0$,
and apply a local perturbation $P_i$ at a distant site $i$. The
perturbed state evolves as
$\rho_P(t) = U(t)P_i\rho_0P_i^\dagger U^\dagger(t)$, with
$U(t)=e^{-iHt}$.
The subsequent growth of $(\Delta B_j)^2_P(t)$ then provides a local
signature of the arrival of the perturbation. Polarized product states
are particularly convenient in this respect, since they can provide a
vanishing initial probe variance together with a spatially localized
perturbation. For short-range Hamiltonians, the response is strongly suppressed outside the effective propagation cone.

The variance alone is most useful when the unperturbed dynamics
provides a sufficiently small or controlled fluctuation background.
For a general initial state, however, $(\Delta B_j)^2$ may already be
nonzero and may evolve substantially even in the absence of the local
perturbation. To isolate the perturbation-induced contribution, we
compare the perturbed evolution $\rho_P(t)$ with the corresponding
unperturbed evolution $\rho_0(t)=U(t)\rho_0U^\dagger(t)$ and define
the differential local variance
\begin{equation}
\mathcal{V}_{i\rightarrow j}(t)
=
\left|
(\Delta B_j)^2_P(t)
-
(\Delta B_j)^2_0(t)
\right|.
\label{eq:differential_variance}
\end{equation}
Here, $(\Delta B_j)^2_0(t)$ and $(\Delta B_j)^2_P(t)$ denote the
variances of the local probe $B_j$ under the unperturbed and
perturbed evolutions, respectively. By subtracting the
unperturbed background, $\mathcal{V}_{i\rightarrow j}(t)$
isolates the perturbation-induced change in local fluctuations.
When the unperturbed variance vanishes identically, as in the
setting considered in Fig.~\ref{scheme}, this expression reduces to $\mathcal{V}_{i\rightarrow j}(t)=(\Delta B_j)^2_P(t)$,
recovering the original local-variance diagnostic.

For $\mathcal{V}_{i\rightarrow j}(t)$ to characterize propagation, the perturbation must leave a spatially separated probe initially unchanged, $\mathcal{V}_{i\rightarrow j}(0)=0$ for $i\neq j$, while acting nontrivially on the initial state, $P_i\rho_0P_i^\dagger\neq\rho_0$. The first condition is automatically satisfied for a local unitary $P_i$ and a strictly local probe $B_j$ with disjoint support, even when $\rho_0$ contains quantum correlations \cite{Beckman2001}. Product initial states therefore provide a simple, although not necessary, realization of these requirements. Among the possible choices of initial states, suitably polarized product states are particularly advantageous for the variance-based diagnostic. In addition to being comparatively simple to prepare in many quantum-simulation platforms \cite{Jurcevic2014,Zhang2017}, polarization along the probe direction ensures a vanishing initial variance, thereby enhancing sensitivity to perturbation-induced changes. If the initial state is also stationary under the unperturbed dynamics, the probe variance remains zero in the absence of the perturbation, providing a background-free signal of its propagation.

For more general initial states, the sensitivity of the diagnostic
depends on the state and its relation to the chosen probe observable.
Thermal states, for example, generally exhibit finite local fluctuations
and reduced polarization, which can weaken the variance response to a propagating perturbation. This limitation becomes extreme at infinite
temperature, where $\rho_\infty=\mathbb{I}/D$ is invariant under any
unitary perturbation, $P_i\rho_\infty P_i^\dagger=\rho_\infty$.
Consequently, the perturbed and unperturbed evolutions coincide and
$\mathcal{V}_{i\rightarrow j}(t)=0$ at all times. Thus, despite its
widespread use in OTOC-based studies \cite{Luitz2017,Khemani2018}, the infinite-temperature state is
insensitive to the deterministic unitary perturbation underlying the
present variance-based protocol.

Although its sensitivity depends on the initial state, the
differential variance admits a general connection to operator
noncommutativity. For a local Pauli probe satisfying
$B_j^2=\mathbb{I}$ and a unitary perturbation $P_i$, we establish
the rigorous bound
\begin{equation}
\mathcal{V}_{i\rightarrow j}(t)
\leq
2\sqrt{C_{BP}^{(i,j)}(t;\rho_0)},
\label{eq:general_variance_otoc}
\end{equation}
where
$C_{BP}^{(i,j)}(t;\rho_0)=
\operatorname{Tr}\!\left\{
\rho_0[B_j(t),P_i]^\dagger[B_j(t),P_i]
\right\}$,
with $B_j(t)=e^{iHt}B_je^{-iHt}$.
This squared commutator quantifies the state-dependent
noncommutativity between the Heisenberg-evolved local probe
$B_j(t)$ and the perturbation $P_i$. The bound establishes that a nonzero differential variance necessarily witnesses operator noncommutativity associated with the applied perturbation. It holds for arbitrary initial states, including correlated and nonstationary states, and is derived in the Supplemental Material \cite{SupplementalMaterial}.

\section{Examples}

To illustrate our approach, we consider a one-dimensional spin-$1/2$
Heisenberg ferromagnet with open boundaries,
\begin{equation}
H=-J\sum_{i=1}^{N-1}
\left(
\sigma_i^x\sigma_{i+1}^x+
\sigma_i^y\sigma_{i+1}^y+
\sigma_i^z\sigma_{i+1}^z
\right),
\label{eq}
\end{equation}
where $N$ is the number of spins, $\sigma_i^\alpha$ are Pauli
operators, and $J>0$ is the ferromagnetic coupling. Despite its
interacting nature, the model admits fully polarized eigenstates that
are exact product states. We consider $N=20$ and initialize the system
in the fully polarized eigenstate
$\ket{\psi_0}=\ket{\uparrow\uparrow\cdots\uparrow}$, where $\ket{\uparrow}$ denotes the eigenstate of $\sigma^z$ with eigenvalue $+1$. At $t=0$, we
locally perturb the first spin by applying $P_1=\sigma_1^x$, producing
the state $\ket{\downarrow\uparrow\cdots\uparrow}$. The resulting state
is no longer stationary, and the initially localized excitation
subsequently propagates through the chain. We monitor this propagation
through the local variance $(\Delta\sigma_j^z)^2(t)$, which initially
vanishes at every site and becomes finite as the influence of the
perturbation reaches the probe site $j$.

\begin{figure}
\includegraphics[width=0.48\textwidth]{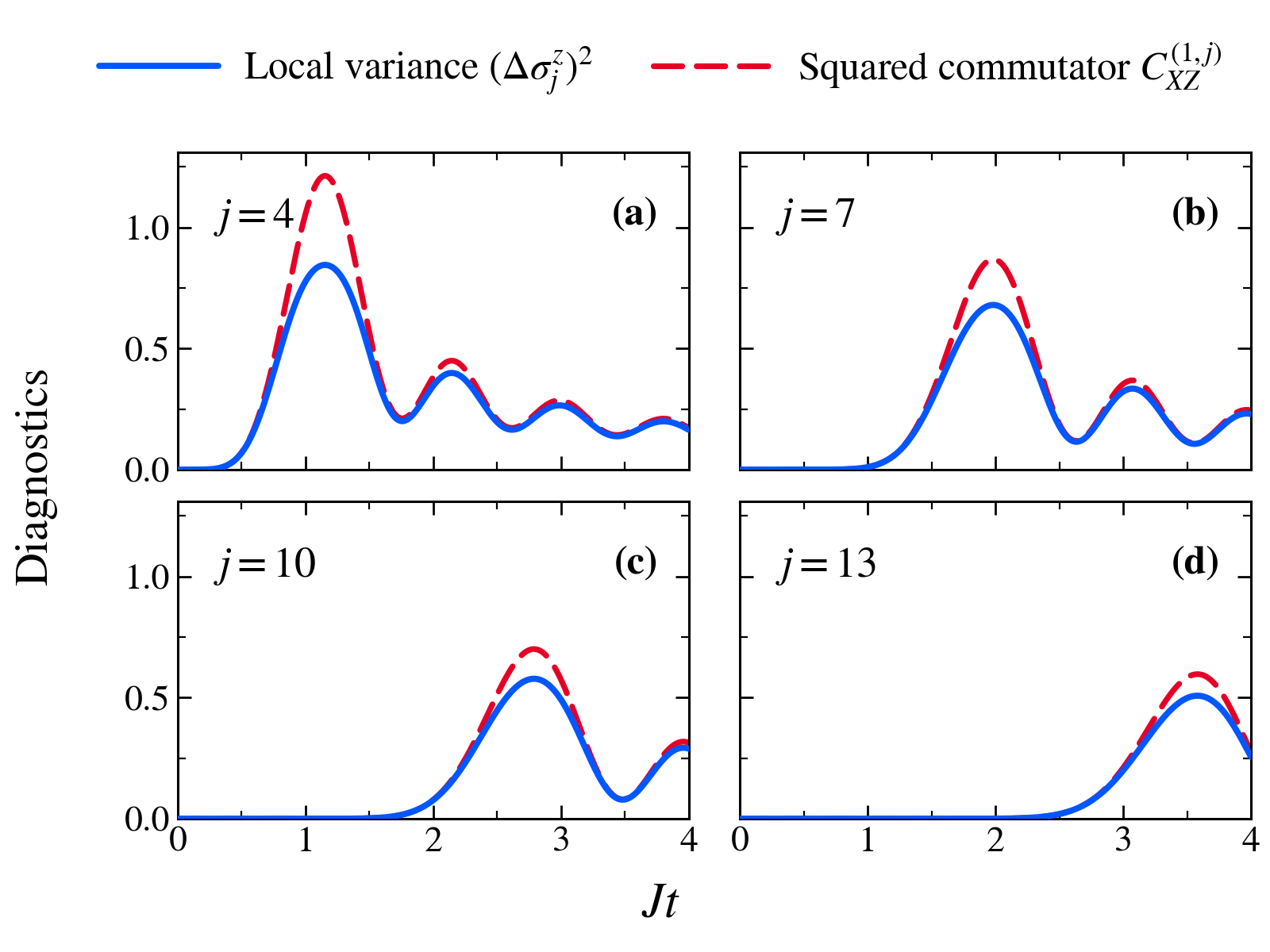}
\caption{
Local variance $(\Delta\sigma_j^z)^2(t)$ (solid blue) and
squared-commutator OTOC $C_{XZ}^{(1,j)}(t)$ (dashed red) for an open
ferromagnetic Heisenberg chain with $N=20$, following a local spin flip
at site $1$. The panels correspond to probe sites
(a) $j=4$, (b) $j=7$, (c) $j=10$, and (d) $j=13$.
The progressive delay in the common onset of the two diagnostics with
increasing distance from the perturbation site tracks the propagation
of the excitation through the chain. We set $\hbar=J=1$, with time
shown as the dimensionless quantity $Jt$.
}
\label{ferro}
\end{figure}

In Fig.~\ref{ferro}, we show the time evolution of
$(\Delta\sigma_j^z)^2(t)$ at several probe sites,
$j=4,7,10,$ and $13$, and compare it with the corresponding
state-dependent squared-commutator $XZ$-OTOC, $C_{XZ}^{(1,j)}(t)
=
\left\langle \psi_0|
[\sigma_1^x(t),\sigma_j^z]^\dagger
[\sigma_1^x(t),\sigma_j^z]
|\psi_0
\right\rangle $, where $\sigma_1^x(t)=e^{iHt}\sigma_1^x e^{-iHt}$. 
Both quantities characterize the propagation associated with
the same local spin-flip operator $\sigma_1^x$, evaluated
relative to the fully polarized initial state. The local variance measures the emergence of fluctuations at the probe site $j$, whereas $C_{XZ}^{(1,j)}(t)$ quantifies the growth of
$\sigma_1^x(t)$ through its failure to commute with the local operator $\sigma_j^z$. As the probe is moved farther from the perturbation site, the onset of both diagnostics is progressively delayed, directly revealing the propagation of the excitation through the chain. Importantly, the local variance and squared commutator exhibit essentially the same onset at each probe site, showing that the propagating front identified by the OTOC can also be detected from
local fluctuations.

For the present model and initial state, this agreement can be
understood analytically. Since the Heisenberg Hamiltonian conserves
the total magnetization, $[H,S_{\mathrm{tot}}^z]=0$, with
$S_{\mathrm{tot}}^z=\frac{1}{2}\sum_i\sigma_i^z$, the dynamics
following a single local spin flip of the fully polarized state
remains confined to the one-excitation sector
\cite{Auerbach1994,Bose2003,Fukuhara2013}. Hence, both diagnostics
are determined by the probability $p_j(t)$ of finding the
excitation at site $j$. In
particular, one finds
$C_{XZ}^{(1,j)}(t)=4p_j(t)$, and
$(\Delta\sigma_j^z)^2(t)=4p_j(t)[1-p_j(t)]$, as shown in the Supplemental  \cite{SupplementalMaterial}. As a result, the two
quantities obey the exact relation
\begin{equation}
(\Delta\sigma_j^z)^2(t)
=
C_{XZ}^{(1,j)}(t)
\left[
1-\frac{C_{XZ}^{(1,j)}(t)}{4}
\right].
\label{eq:exact_variance_otoc}
\end{equation}
At the leading edge of the propagating excitation, where
$p_j(t)\ll1$, this reduces to
$(\Delta\sigma_j^z)^2(t)\simeq C_{XZ}^{(1,j)}(t)$, with deviations
appearing only at second order in $p_j(t)$. This behavior is directly
reflected in Fig.~\ref{ferro}: although the two quantities may differ
behind the propagating front, their leading edges closely coincide,
and their common onset shifts to later times as the probe distance
increases. The local variance therefore reproduces, to leading order,
the propagation front detected by the squared-commutator OTOC while
requiring only local expectation values. We emphasize that the exact
relation in Eq.~(\ref{eq:exact_variance_otoc}) is specific to the one-excitation dynamics and polarized reference state considered here.

We now illustrate the differential local variance in an open
Ising chain governed by the Hamiltonian
\begin{equation}
H=-\sum_{i=1}^{N-1}\sigma_i^z\sigma_{i+1}^z
+g\sum_{i=1}^{N}\sigma_i^x
+h\sum_{i=1}^{N}\sigma_i^z,
\label{eq:ising_model}
\end{equation}
where the nearest-neighbor coupling is set to unity.
For $h=0$, the model reduces to the integrable
transverse-field Ising chain, whereas a finite
longitudinal field $h$ generally breaks integrability
when $g\neq 0$. We consider $N=9$ spins with $g=1$
and $h=0.5$, corresponding to the nonintegrable
regime, initially prepared in
$\ket{\psi_0}=\ket{+}^{\otimes N}$, where
$\sigma^x\ket{+}=\ket{+}$. A local perturbation
is introduced at site $1$ through
$R_z(\theta)=\exp(-i\theta\sigma_1^z/2)$,
with $\theta=\pi$ or $\pi/2$. Its propagation
is monitored through the differential local variance
$\mathcal{V}_{\theta}^{(1,j)}(t)
=
\left|
(\Delta\sigma_j^x)^2_{\theta}(t)
-
(\Delta\sigma_j^x)^2_{0}(t)
\right|$,
where
$(\Delta\sigma_j^x)^2_{\theta}(t)
=1-\langle\sigma_j^x\rangle_{\theta,t}^{2}$
is evaluated in
$\ket{\psi_{\theta}(t)}
=e^{-iHt}R_z(\theta)\ket{\psi_0}$,
while the subscript $0$ denotes the unperturbed
evolution
$\ket{\psi(t)}=e^{-iHt}\ket{\psi_0}$.
We also calculate the {\it normalized} squared
commutators
$C_{\alpha X}^{(1,j)}(t)
=\frac14\bra{\psi_0}
[\sigma_1^\alpha(t),\sigma_j^x]^\dagger
[\sigma_1^\alpha(t),\sigma_j^x]
\ket{\psi_0}$,
with $\alpha=y,z$ and
$\sigma_1^\alpha(t)
=e^{iHt}\sigma_1^\alpha e^{-iHt}$. Here, the OTOCs evolve the operator at the perturbation
site, whereas the bound in Eq.~(7) involves the evolved probe through $[\sigma_j^x(t),R_z(\theta)]$. These squared commutators need not coincide for the nonstationary initial state considered here. The comparison below therefore examines qualitative propagation signatures rather than directly testing Eq.~(\ref{eq:general_variance_otoc}).

\begin{figure}[t]
\centering
\includegraphics[width=0.48\textwidth]{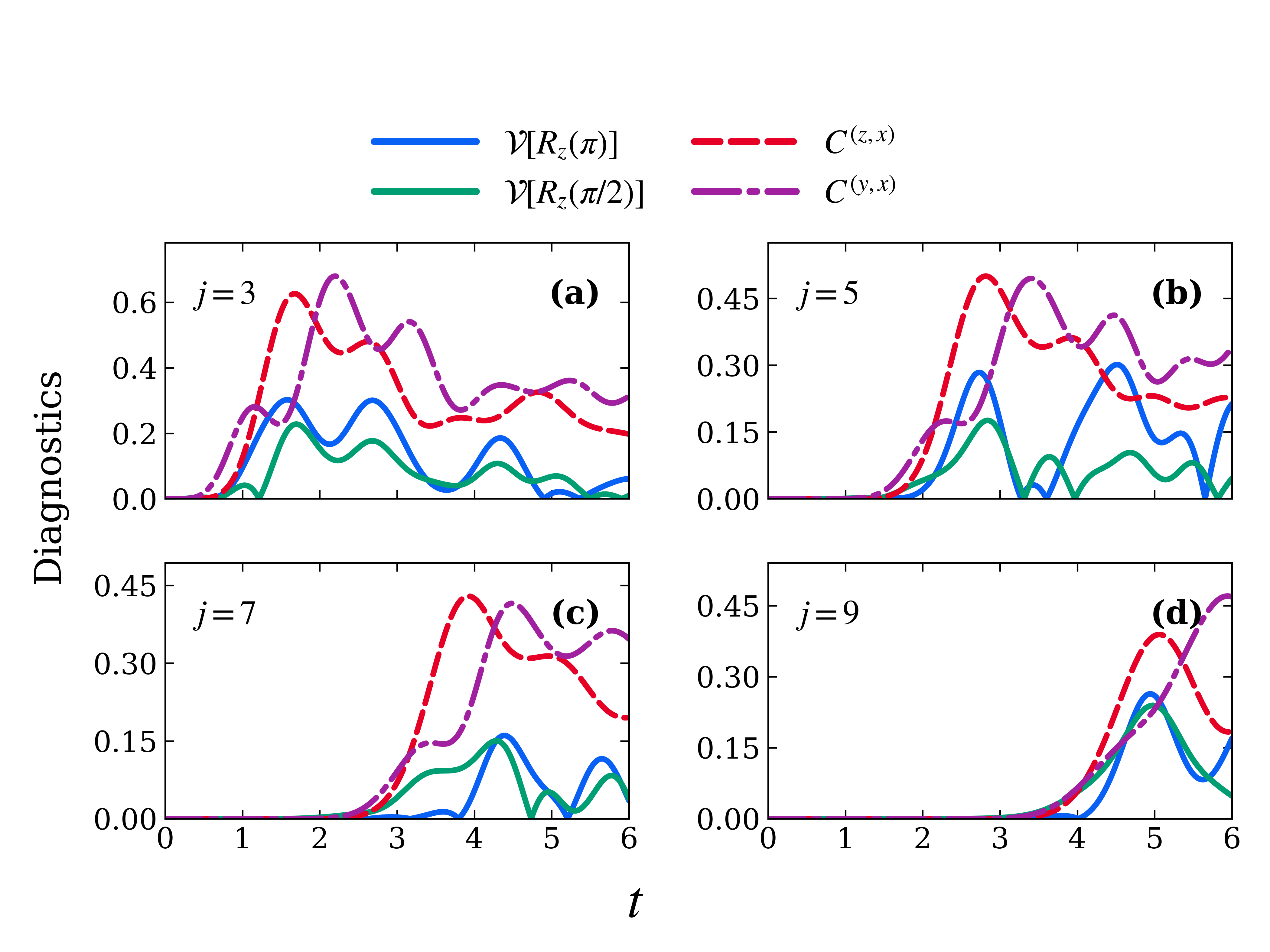}
\caption{
Differential local variances
$\mathcal{V}_{\theta}^{(1,j)}(t)$ and normalized
squared commutators $C_{\alpha X}^{(1,j)}(t)$
in an open Ising chain with $N=9$, $g=1$,
and $h=0.5$, in the nonintegrable regime.
The initial state is $\ket{+}^{\otimes N}$.
Blue and green solid curves represent the
differential variances induced by local Bloch-sphere
rotations about the $z$ axis at site $1$, with
angles $\theta=\pi$ and $\theta=\pi/2$,
respectively.
Red dashed and purple dash-dotted curves denote
$C_{ZX}^{(1,j)}(t)$ and $C_{YX}^{(1,j)}(t)$,
respectively.
Panels (a)--(d) correspond to probe sites
$j=3,5,7,9$.
}
\label{Ising}
\end{figure}

Figure~\ref{Ising} presents the evolution of these
diagnostics for probe sites $j=3,5,7,9$,
corresponding to increasing distances from the
perturbation site. Both the differential variances
and squared commutators exhibit a progressively
delayed onset as the separation increases,
consistent with the propagation of local
perturbations through the chain. Their onset
timescales display qualitatively similar
distance dependence, although their subsequent
temporal behavior is generally distinct. Importantly, differential variances and squared
commutators are different physical quantities,
with distinct operational definitions.
Consequently, their numerical amplitudes should not be directly compared. The comparison
in Fig.~\ref{Ising} concerns their qualitative
dynamical features, particularly the emergence
and propagation of local signals.

These results indicate that differential local
variance provides a state-dependent diagnostic
of perturbation propagation complementary to
conventional OTOCs. While the former quantifies
the change in local fluctuations induced by a
specific perturbation, the latter characterizes
the noncommutativity between a Heisenberg-evolved
operator and a distant observable. In particular,
although $R_z(\pi)$ is equivalent to $\sigma_1^z$ up to a global phase, the associated
differential variance is not identical to
$C_{ZX}^{(1,j)}(t)$. The qualitative correspondence
between their propagation features therefore
does not imply an equivalence between the two
diagnostics. Results for the integrable limit
$h=0$ and for the nonintegrable case $h=1$
are presented in the Supplemental Material \cite{SupplementalMaterial},
extending the analysis across different
dynamical regimes.
\\

\section{Discussion and conclusions}

We have introduced a local-fluctuation approach to
probing operator spreading in interacting quantum
many-body systems. By monitoring the change in the
variance of a local observable following a spatially
separated perturbation, the method identifies
propagation signatures using only forward time
evolution and local measurements. Motivated by the Heisenberg uncertainty relation, we established a
rigorous bound connecting the differential local
variance to operator noncommutativity, providing
a theoretical foundation for the proposed diagnostic.
For a ferromagnetic Heisenberg chain in the one-excitation regime, we derived an exact analytical relation between the local variance and a
squared-commutator OTOC, demonstrating their
agreement at the leading propagation front. For an Ising chain, we found qualitatively consistent propagation timescales for the differential variance and squared commutators in both integrable and nonintegrable regimes.

These findings do not imply an equivalence between the diagnostics. While squared commutators quantify
operator noncommutativity, differential local variance measures the perturbation-induced change in local fluctuations for a given initial state and probe observable. Their numerical magnitudes are therefore not directly comparable, and their correspondence concerns the qualitative features of perturbation propagation. Moreover, the variance-based diagnostic may become insensitive for certain initial states, notably the infinite-temperature state, for which a unitary perturbation leaves the density matrix unchanged. Identifying optimal initial states and local probes, and investigating the effects
of noise and decoherence, are important directions for future work. Our results establish local quantum fluctuations as a complementary and
experimentally accessible means of probing operator spreading without requiring explicit many-body time reversal.

\section*{Acknowledgements}

The author acknowledges support from Conselho Nacional de Desenvolvimento Cient{\'i}fico e Tecnol{\'o}gico (CNPq, Grant No. 318905/2025-6).

\section*{Data availability}

There are no publicly available research data or software supporting this manuscript. Requests for further information or data should be sent to the author.

\clearpage
\onecolumngrid

\begin{center}
    \large\bfseries
    Supplemental Material for:\\
    Probing Operator Spreading with Local Quantum Fluctuations
\end{center}

\setcounter{equation}{0}
\setcounter{figure}{0}
\setcounter{table}{0}

\renewcommand{\theequation}{S\arabic{equation}}
\renewcommand{\thefigure}{S\arabic{figure}}
\renewcommand{\thetable}{S\arabic{table}}

\medskip

\noindent
This Supplemental Material provides additional analytical derivations
and numerical results supporting the findings of the main text.
We first derive the spectral representation of the response function
and establish a rigorous bound relating the differential local variance
to a state-dependent squared commutator. We then obtain an exact
relation between the local variance and the XZ squared-commutator OTOC
in the one-excitation sector of the Heisenberg chain. Finally, we
present additional results for the Ising chain in both integrable
($h=0$) and nonintegrable ($h=1$) regimes, further examining the
propagation signatures captured by the proposed diagnostic.

\medskip

\section{Calculation of the response function $\chi(t)$}

We consider a many-body system with Hamiltonian $H$, eigenstates $\{|n\rangle\}$, and eigenvalues $E_n$, i.e.\ $H|n\rangle=E_n|n\rangle$. For two (generally local) operators $A$ and $B$, we define the Heisenberg-picture operator
\begin{equation}
  A(t) \equiv e^{iHt} A e^{-iHt}.
\end{equation}
Given a density matrix $\rho$, the \emph{response function} is \cite{Kubo1957}
\begin{equation}
\label{response}
  \chi(t) \equiv -\,i\,\langle [A(t),B]\rangle \equiv -\,i\,\mathrm{Tr}\!\left(\rho\,[A(t),B]\right)
  \;=\; -\,i\left[\mathrm{Tr}\!\left(\rho\,A(t)B\right) - \mathrm{Tr}\!\left(\rho\,B A(t)\right)\right].
\end{equation}

In what follows, we calculate both terms separately. The first one is 
\begin{eqnarray}
\mathrm{Tr}\!\left(\rho\,A(t)B\right) = \mathrm{Tr}\!\left(\rho\, e^{iHt} A e^{-iHt}B\right). 
\end{eqnarray}
If we assume that $\rho$ is diagonal in the energy basis, i.e.\ $[\rho,H]=0$, which is the case of the stationary eigenstates of the Hamiltonian, we can write $\rho=\sum_{a}\rho_{aa}|a\rangle\langle a|$. With this, we obtain
\begin{eqnarray}
\mathrm{Tr}\!\left(\rho\,A(t)B\right) &=& \sum_k \bra{k}\left( \sum_{a}\rho_{aa}|a\rangle\langle a| \sum_{m}e^{iE_mt} \ket{m}\bra{m}A\sum_{n}e^{-iE_nt} \ket{n}\bra{n}B\right)\ket{k} \nonumber \\
&=&\sum_{k,a,m,n}\rho_{aa} \braket{k|a} \braket{a|m} \braket{m|A|n}\braket{n|B|k}e^{i(E_m-E_n)t}. 
\end{eqnarray}
Using $\braket{k|a} = \delta_{k,a}$ and $\braket{a|m} = \delta_{a,m}$ reduces the above expression to 
\begin{eqnarray}
\label{first}
\mathrm{Tr}\!\left(\rho\,A(t)B\right) 
&=&\sum_{m,n}\rho_{mm} A_{mn}B_{nm} e^{i(E_m-E_n)t}, 
\end{eqnarray}
where $A_{mn}\equiv \langle m|A|n\rangle$, and similarly $B_{nm}\equiv \langle n|B|m\rangle$.

The second term of Eq.~(\ref{response}) is calculated as follows:
\begin{eqnarray}
\mathrm{Tr}\!\left(\rho\,BA(t)\right) &=& \mathrm{Tr}\!\left(\rho\, B e^{iHt} A e^{-iHt}\right) \nonumber \\
&=& \sum_k \bra{k}\left( \sum_{a}\rho_{aa}|a\rangle\langle a| B\sum_{m}e^{iE_mt} \ket{m}\bra{m}A\sum_{n}e^{-iE_nt} \ket{n}\bra{n}\right)\ket{k} \nonumber \\
&=&\sum_{k,a,m,n}\rho_{aa} \braket{k|a} \braket{a|B|m} \braket{m|A|n}\braket{n|k}e^{i(E_m-E_n)t},
\end{eqnarray}
which provides
\begin{eqnarray}
\label{second}
\mathrm{Tr}\!\left(\rho\,BA(t)\right) 
&=&\sum_{m,n}\rho_{nn} B_{nm}A_{mn} e^{i(E_m-E_n)t}. 
\end{eqnarray}
By substitution of Eqs.~(\ref{first}) and~(\ref{second}) into Eq.~(\ref{response}), we finally get 
\begin{equation}
\chi(t)= -i \sum_{m,n} \big(\rho_{mm}-\rho_{nn}\big)\, A_{mn} B_{nm}\, e^{i(E_m-E_n)t}.
\label{oscillations}
\end{equation}

Equation~\eqref{oscillations} shows that $\chi(t)$ is a \emph{signed} superposition of oscillatory contributions with phases $e^{i(E_m-E_n)t}$. Even if the support of $A(t)$ grows and overlaps with $B$, destructive interference among these frequencies can render $\chi(t)$ small or even zero at various times. This sensitivity to phase cancellations underlies the limitations of $\chi(t)$ as a monotone diagnostic of operator spreading and motivates the use of squared-commutator OTOCs.

\section{Bound between the differential variance and the squared-commutator OTOC}

We derive a bound relating the perturbation-induced change in the local variance to a state-dependent squared commutator involving the applied perturbation. Consider an arbitrary initial density operator $\rho_0$ and a local unitary perturbation $P_i$ applied at site $i$. The unperturbed and perturbed states evolve as
\begin{equation}
\rho_0(t)=U(t)\rho_0U^\dagger(t),
\qquad
\rho_P(t)=U(t)P_i\rho_0P_i^\dagger U^\dagger(t),
\end{equation}
where $U(t)=e^{-iHt}$. We denote the corresponding variances of a local probe observable $B_j$ by $(\Delta B_j)^2_0(t)$ and $(\Delta B_j)^2_P(t)$, respectively.

For a Pauli probe satisfying $B_j^2=\mathbb{I}$, these variances take the form
\begin{equation}
(\Delta B_j)^2_0(t)=1-m^2(t),
\qquad
(\Delta B_j)^2_P(t)=1-m_P^2(t),
\end{equation}
where
\begin{equation}
m(t)=\operatorname{Tr}[\rho_0(t)B_j],
\qquad
m_P(t)=\operatorname{Tr}[\rho_P(t)B_j].
\end{equation}
The differential variance introduced in the main text can therefore be written as
\begin{align}
\mathcal{V}_{i\rightarrow j}(t)
&=
\left|
(\Delta B_j)^2_P(t)-(\Delta B_j)^2_0(t)
\right|
\nonumber\\
&=
\left|m_P^2(t)-m^2(t)\right|
\nonumber\\
&=
|m_P(t)-m(t)|\,|m_P(t)+m(t)|.
\label{eq:supp_var_difference}
\end{align}
Since $B_j$ is a Pauli observable with eigenvalues $\pm 1$,
its expectation values satisfy $|m(t)|\leq 1$ and
$|m_P(t)|\leq 1$. Using the triangle inequality, we obtain
\begin{align}
\mathcal{V}_{i\rightarrow j}(t)
&=|m_P(t)-m(t)|\,|m_P(t)+m(t)|
\nonumber\\
&\leq |m_P(t)-m(t)|
\bigl(|m_P(t)|+|m(t)|\bigr)
\nonumber\\
&\leq 2|m_P(t)-m(t)|.
\label{eq:supp_var_first_bound}
\end{align}

To connect the variance response to operator noncommutativity, we introduce the Heisenberg-picture probe
\begin{equation}
B_j(t)=U^\dagger(t)B_jU(t).
\end{equation}
The difference between the perturbed and unperturbed expectation values becomes
\begin{align}
m_P(t)-m(t)
&=
\operatorname{Tr}\!\left[
\rho_0\left(
P_i^\dagger B_j(t)P_i-B_j(t)
\right)
\right]
\nonumber\\
&=
\operatorname{Tr}\!\left[
\rho_0 P_i^\dagger[B_j(t),P_i]
\right],
\label{eq:supp_expectation_difference}
\end{align}
where the second equality follows from the unitarity of $P_i$.

Defining $X(t)=[B_j(t),P_i]$ and applying the Cauchy--Schwarz inequality with respect to $\rho_0$, we obtain
\begin{align}
|m_P(t)-m(t)|^2
&=
\left|
\operatorname{Tr}[\rho_0P_i^\dagger X(t)]
\right|^2
\nonumber\\
&\leq
\operatorname{Tr}[\rho_0P_i^\dagger P_i]\,
\operatorname{Tr}[\rho_0X^\dagger(t)X(t)]
\nonumber\\
&=
\operatorname{Tr}[\rho_0X^\dagger(t)X(t)],
\end{align}
where we used $P_i^\dagger P_i=\mathbb{I}$ and $\operatorname{Tr}\rho_0=1$.

We define the perturbation-associated state-dependent squared commutator as
\begin{equation}
C_{BP}^{(i,j)}(t;\rho_0)
=
\left\langle
[B_j(t),P_i]^\dagger
[B_j(t),P_i]
\right\rangle_{\rho_0},
\label{eq:supp_otoc_general}
\end{equation}
where $\langle O\rangle_{\rho_0}=\operatorname{Tr}(\rho_0O)$. The preceding inequality then yields
\begin{equation}
|m_P(t)-m(t)|
\leq
\sqrt{C_{BP}^{(i,j)}(t;\rho_0)}.
\label{eq:supp_expectation_otoc}
\end{equation}
Combining this result with Eq.~\eqref{eq:supp_var_first_bound}, we obtain
\begin{equation}
\mathcal{V}_{i\rightarrow j}(t)
\leq
2\sqrt{C_{BP}^{(i,j)}(t;\rho_0)}.
\label{eq:supp_variance_otoc_bound}
\end{equation}

This bound holds for arbitrary initial density operators, including pure, mixed, correlated, and nonstationary states, provided that $P_i$ is unitary and $B_j^2=\mathbb{I}$. In particular,
\begin{equation}
\mathcal{V}_{i\rightarrow j}(t)>0
\quad\Longrightarrow\quad
C_{BP}^{(i,j)}(t;\rho_0)>0.
\end{equation}
A nonzero differential variance therefore witnesses noncommutativity between the Heisenberg-evolved probe $B_j(t)$ and the applied perturbation $P_i$. The converse does not generally hold, since a nonzero squared commutator need not produce a change in the variance of the chosen probe.

\section{Exact relation between the local variance and the $XZ$-OTOC in the one-excitation sector}
\label{sec:variance_otoc}

We consider the open spin-$1/2$ ferromagnetic Heisenberg chain
introduced in the main text,
\begin{equation}
H=-J\sum_{i=1}^{N-1}
\left(
\sigma_i^x\sigma_{i+1}^x+
\sigma_i^y\sigma_{i+1}^y+
\sigma_i^z\sigma_{i+1}^z
\right),
\label{eq:H_supp}
\end{equation}
where $N$ is the number of spins, and the fully polarized reference
state
\begin{equation}
\ket{\psi_0}
=
\ket{\uparrow\uparrow\cdots\uparrow}.
\label{eq:psi0_supp}
\end{equation}
The latter is an eigenstate of the Hamiltonian,
\begin{equation}
H\ket{\psi_0}
=
E_0\ket{\psi_0},
\qquad
E_0=-(N-1)J.
\label{eq:E0}
\end{equation}

To generate nontrivial dynamics, at $t=0$ we apply the local
perturbation $P_1=\sigma_1^x$, which flips the first spin. The
resulting state is
\begin{equation}
\ket{\psi(0)}
=
\sigma_1^x\ket{\psi_0}
=
\ket{\downarrow\uparrow\cdots\uparrow}
\equiv
\ket{1}.
\label{eq:initial_magnon}
\end{equation}
Since the Heisenberg Hamiltonian conserves the total magnetization,
the subsequent evolution is restricted to the one-excitation sector \cite{Auerbach1994}.
Introducing the basis $\{\ket{j}\}_{j=1}^{N}$, where $\ket{j}$
denotes the state in which the only down spin is located at site
$j$, the evolved state can be written as
\begin{equation}
\ket{\psi(t)}
=
e^{-iHt}\ket{1}
=
\sum_{j=1}^{N}a_j(t)\ket{j},
\label{eq:psi_t_magnon}
\end{equation}
where
\begin{equation}
p_j(t)=|a_j(t)|^2
\label{eq:pj}
\end{equation}
is the probability of finding the spin excitation at site $j$.

Within the one-excitation sector, the Heisenberg
Hamiltonian maps exactly onto a nearest-neighbor
tight-binding model. Using the raising and lowering
operators $\sigma_i^\pm=(\sigma_i^x\pm i\sigma_i^y)/2$,
the transverse exchange interaction can be expressed as
\begin{equation}
\sigma_i^x\sigma_{i+1}^x+
\sigma_i^y\sigma_{i+1}^y
=
2\left(
\sigma_i^+\sigma_{i+1}^-+
\sigma_i^-\sigma_{i+1}^+
\right).
\end{equation}
These terms exchange neighboring up and down spins,
allowing the single spin-flip excitation to hop
between adjacent sites with amplitude $-2J$.
The longitudinal interaction,
$\sigma_i^z\sigma_{i+1}^z$, contributes diagonal
energy shifts. Relative to the energy $E_0$ of
the fully polarized state, these shifts are $4J$
for an excitation in the bulk and $2J$ at either
boundary, reflecting the number of neighboring
bonds affected by the spin flip. Consequently,
for $2\leq j\leq N-1$,
\begin{equation}
(H-E_0)\ket{j}
=
4J\ket{j}
-2J\ket{j-1}
-2J\ket{j+1},
\label{eq:H_bulk}
\end{equation}
whereas at the boundaries,
\begin{align}
(H-E_0)\ket{1}
&=
2J\ket{1}-2J\ket{2},
\\
(H-E_0)\ket{N}
&=
2J\ket{N}-2J\ket{N-1}.
\label{eq:H_boundaries}
\end{align}
Thus, the dynamics of the initially localized
spin flip can be determined by diagonalizing
an $N\times N$ effective Hamiltonian instead of
the full $2^N$-dimensional many-body Hamiltonian.

\subsection{Local variance}

We characterize the arrival of the excitation at site $j$ through
the variance of the local Pauli operator $\sigma_j^z$, denoted by
$(\Delta\sigma_j^z)^2(t)$:
\begin{equation}
(\Delta\sigma_j^z)^2(t)
=
\mel{\psi(t)}{(\sigma_j^z)^2}{\psi(t)}
-
\mel{\psi(t)}{\sigma_j^z}{\psi(t)}^2.
\label{eq:variance_def}
\end{equation}
Since $(\sigma_j^z)^2=\mathbb{I}$, this reduces to
\begin{equation}
(\Delta\sigma_j^z)^2(t)
=
1-
\mel{\psi(t)}{\sigma_j^z}{\psi(t)}^2.
\label{eq:variance_sigma}
\end{equation}

In the one-excitation sector, the spin at site $j$ is down with
probability $p_j(t)$ and up with probability $1-p_j(t)$.
Consequently,
\begin{equation}
\mel{\psi(t)}{\sigma_j^z}{\psi(t)}
=
[1-p_j(t)]-p_j(t)
=
1-2p_j(t),
\label{eq:sigmaz_expectation}
\end{equation}
which gives the exact relation
\begin{equation}
(\Delta\sigma_j^z)^2(t)
=
4p_j(t)[1-p_j(t)].
\label{eq:variance_p}
\end{equation}

At $t=0$, one has $p_1(0)=1$ and $p_{j>1}(0)=0$, so that
$(\Delta\sigma_j^z)^2(0)=0$ at every site. For a spatially
separated probe $j>1$, a nonzero local variance develops as the
propagating excitation acquires a finite probability of occupying
that site.

\subsection{Squared-commutator $XZ$-OTOC}

We compare the local variance with the state-dependent $XZ$
squared commutator
\begin{equation}
C_{XZ}^{(1,j)}(t)
=
\bra{\psi_0}
[\sigma_1^x(t),\sigma_j^z]^\dagger
[\sigma_1^x(t),\sigma_j^z]
\ket{\psi_0},
\label{eq:otoc_def}
\end{equation}
where
\begin{equation}
\sigma_1^x(t)
=
e^{iHt}\sigma_1^x e^{-iHt}.
\label{eq:sigmax_heisenberg}
\end{equation}

Because the fully polarized state is an eigenstate of $H$, the
action of the time-evolved operator on the reference state is
\begin{align}
\sigma_1^x(t)\ket{\psi_0}
&=
e^{iHt}\sigma_1^x e^{-iHt}\ket{\psi_0}
\nonumber\\
&=
e^{-iE_0t}e^{iHt}\sigma_1^x\ket{\psi_0}
\nonumber\\
&=
e^{-iE_0t}e^{iHt}\ket{1}.
\label{eq:operator_state_relation}
\end{align}

Since the Hamiltonian has real matrix elements
in the one-excitation basis,
$\langle j|H|k\rangle\in\mathbb{R}$,
the corresponding time-evolution operators satisfy
$e^{iHt}=(e^{-iHt})^*$ in this basis.
Consequently, using Eq.~\eqref{eq:psi_t_magnon},
we obtain
\begin{equation}
\sigma_1^x(t)\ket{\psi_0}
=
e^{-iE_0t}
\sum_{\ell=1}^{N}a_\ell^*(t)\ket{\ell}.
\label{eq:operator_state_expansion}
\end{equation}
The complex conjugation does not affect the probabilities
$|a_\ell(t)|^2$ entering the squared commutator.

Furthermore,
\begin{equation}
\sigma_j^z\ket{\psi_0}
=
\ket{\psi_0}.
\label{eq:sigmaz_psi0}
\end{equation}
The commutator acting on the reference state can therefore be
written as
\begin{align}
[\sigma_1^x(t),\sigma_j^z]\ket{\psi_0}
&=
\sigma_1^x(t)\ket{\psi_0}
-
\sigma_j^z\sigma_1^x(t)\ket{\psi_0}
\nonumber\\
&=
(\mathbb{I}-\sigma_j^z)
\sigma_1^x(t)\ket{\psi_0}.
\label{eq:commutator_action}
\end{align}

For a one-excitation basis state $\ket{\ell}$,
\begin{equation}
(\mathbb{I}-\sigma_j^z)\ket{\ell}
=
2\delta_{\ell j}\ket{j}.
\label{eq:projector_action}
\end{equation}
It follows that
\begin{equation}
[\sigma_1^x(t),\sigma_j^z]\ket{\psi_0}
=
2e^{-iE_0t}a_j^*(t)\ket{j}.
\label{eq:commutator_final}
\end{equation}

Taking the norm squared yields
\begin{equation}
C_{XZ}^{(1,j)}(t)
=
4|a_j(t)|^2
=
4p_j(t).
\label{eq:otoc_p}
\end{equation}

Thus, for the fully polarized reference state considered here,
both $(\Delta\sigma_j^z)^2(t)$ and the state-dependent squared
commutator are determined by the same single-excitation propagation
probability $p_j(t)$. Combining Eqs.~\eqref{eq:variance_p} and
\eqref{eq:otoc_p}, we obtain the exact relation
\begin{equation}
(\Delta\sigma_j^z)^2(t)
=
C_{XZ}^{(1,j)}(t)
\left[
1-\frac{C_{XZ}^{(1,j)}(t)}{4}
\right].
\label{eq:variance_otoc_relation}
\end{equation}

This identity follows from the restricted one-excitation dynamics,
the fully polarized reference state, and the specific choices of
perturbation and probe employed here. It should therefore not be
interpreted as a general identity between local variances and
OTOCs.

\subsection{Behavior at the propagation front}

The connection between the two diagnostics becomes particularly
transparent at the leading edge of the propagating excitation \cite{LiebRobinson1972}. At a spatially separated probe site $j$, the occupation probability
is small as the leading edge first reaches the probe,
$p_j(t)\ll1$. In this regime,
\begin{equation}
C_{XZ}^{(1,j)}(t)
=
4p_j(t),
\label{eq:otoc_small_p}
\end{equation}
whereas
\begin{align}
(\Delta\sigma_j^z)^2(t)
&=
4p_j(t)[1-p_j(t)]
\nonumber\\
&=
4p_j(t)-4p_j^2(t).
\label{eq:variance_small_p}
\end{align}
Consequently,
\begin{equation}
(\Delta\sigma_j^z)^2(t)
\simeq
C_{XZ}^{(1,j)}(t),
\qquad
p_j(t)\ll1.
\label{eq:variance_otoc_leading}
\end{equation}

More precisely, their difference is
\begin{align}
C_{XZ}^{(1,j)}(t)
-
(\Delta\sigma_j^z)^2(t)
=
4p_j^2(t)
=
\frac{\left[C_{XZ}^{(1,j)}(t)\right]^2}{4}.
\label{eq:otoc_variance_difference}
\end{align}
The difference between the two diagnostics is therefore second
order in the small occupation probability. Accordingly, the local
variance and squared commutator have the same leading behavior as
the propagating front first reaches a distant probe, whereas
deviations become appreciable behind the front when $p_j(t)$ is no
longer small.

This behavior underlies the comparison presented in the main text.
For $N=20$ and $J=1$, with the perturbation applied at site $1$,
we monitor the two diagnostics at the probe sites $j=4,7,10,$ and $13$ over the interval $0\leq Jt\leq4$ (see Fig. 2 in the main text). As the probe is moved farther from the perturbation site, the onset
of both quantities is progressively delayed, reflecting the
propagation of the initially localized excitation through the
chain. The local variance therefore reproduces, to leading order, the
propagation front detected by the state-dependent squared
commutator while requiring only local expectation values.

\section{Additional results for the Ising spin chain}

To examine the dependence of the proposed diagnostic on
integrability, we extend the analysis of the main text
to the integrable transverse-field Ising chain ($h=0$)
and the nonintegrable regime ($h=1$) \cite{CalabreseCardy2006,Khemani2018}, keeping $J=g=1$
and $N=9$. The initial state, local rotations, and
probe observables are identical to those used in the
main text ($h=0.5$).

\begin{figure}[H]
\centering
\includegraphics[width=0.55\textwidth]
{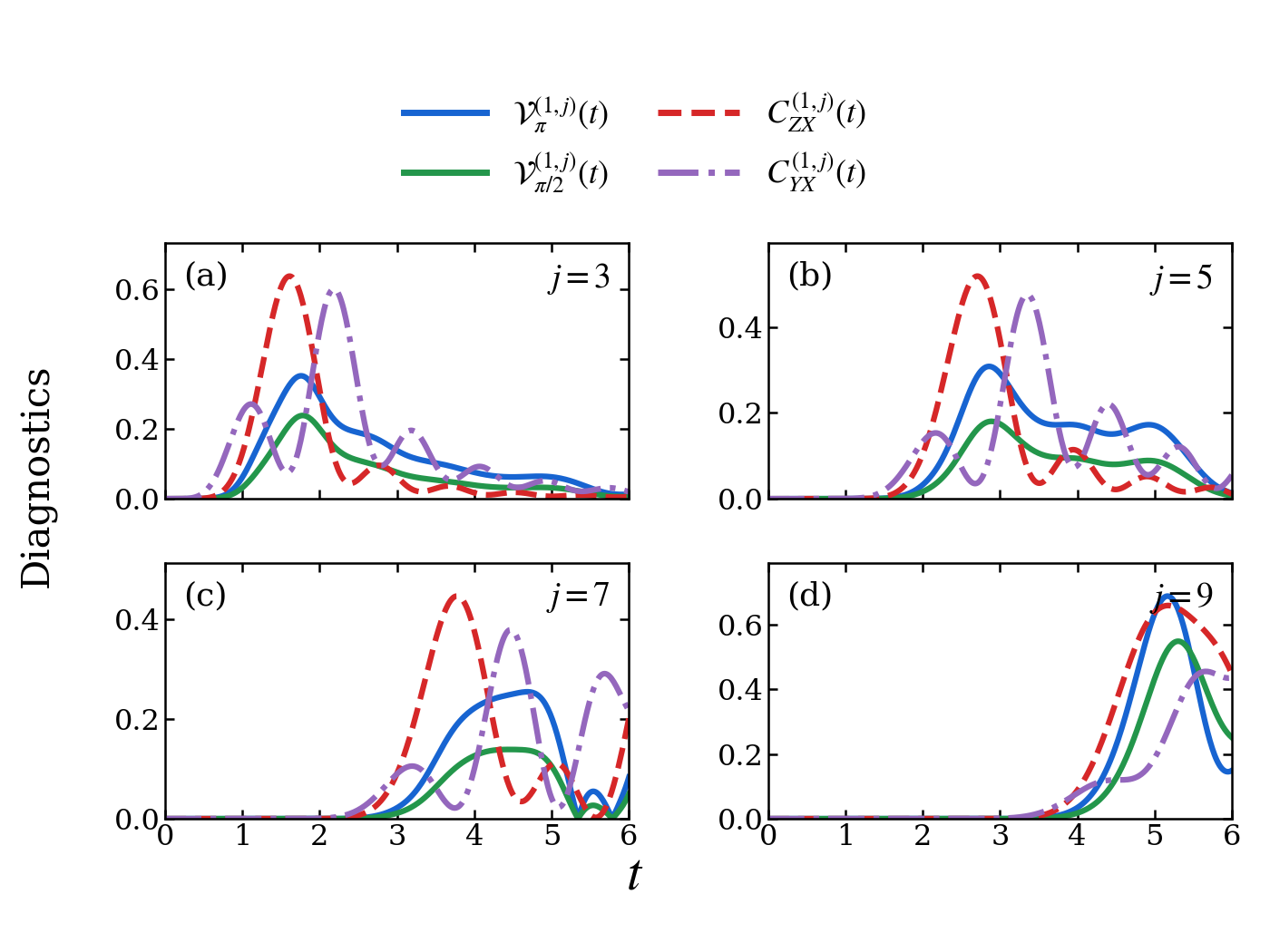}

\caption{
Differential local variance and normalized
squared-commutator OTOCs in the integrable Ising
chain ($h=0$), with $N=9$ and $J=g=1$.
Starting from the initial state
$|\psi_0\rangle=|+\rangle^{\otimes N}$,
local perturbations are applied at site $1$
through the rotation
$R_z(\theta)=e^{-i\theta\sigma_1^z/2}$.
The differential variances of $\sigma_j^x$
for $\theta=\pi$ (blue solid) and
$\theta=\pi/2$ (green solid) are compared
with the normalized squared-commutator OTOCs
$C_{ZX}^{(1,j)}(t)$ (red dashed) and
$C_{YX}^{(1,j)}(t)$ (purple dash-dotted)
at (a) $j=3$, (b) $j=5$, (c) $j=7$,
and (d) $j=9$.
}
\label{Fig1sm}
\end{figure}

\begin{figure}[H]
\centering
\includegraphics[width=0.55\textwidth]
{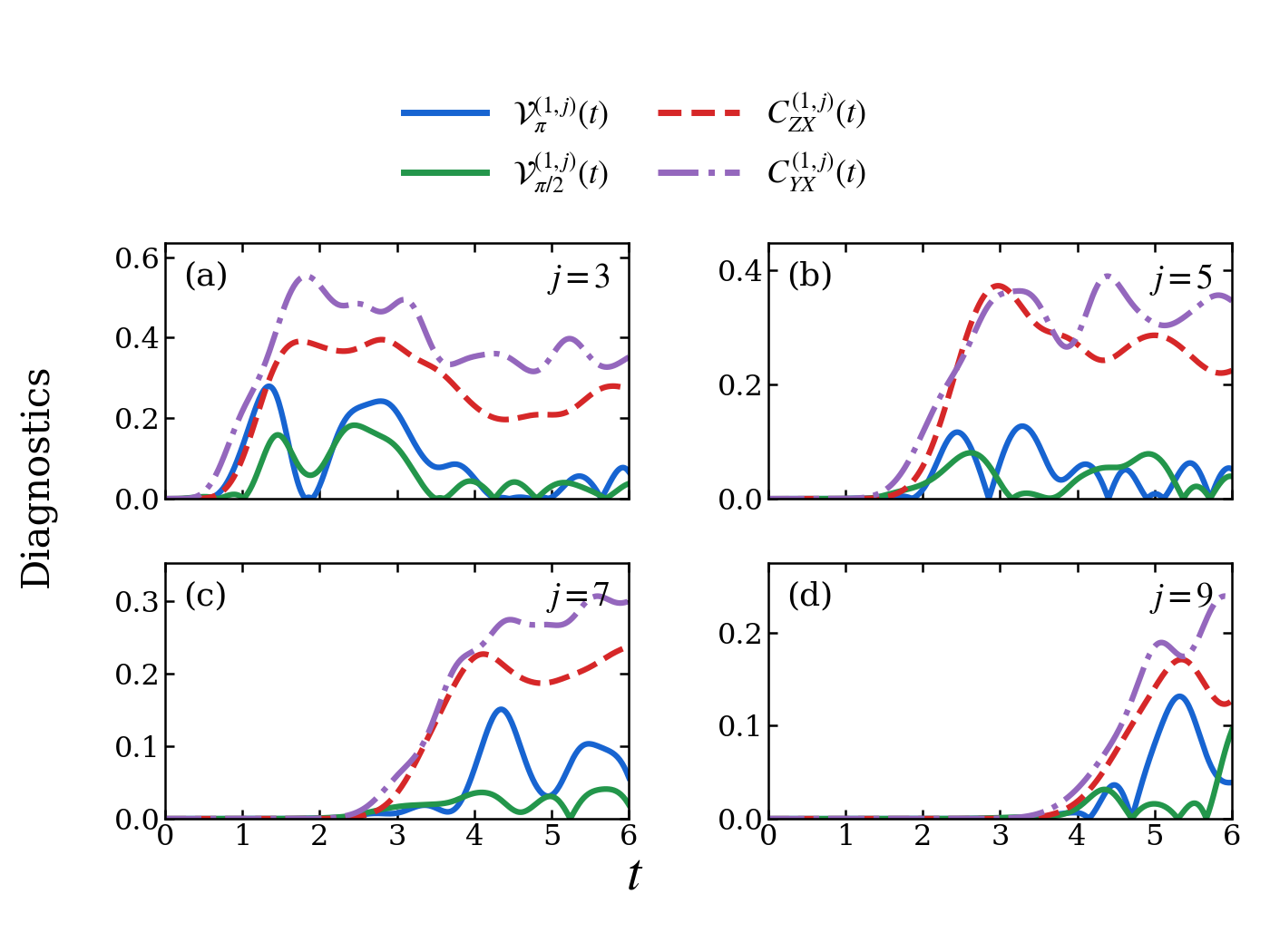}
\caption{
Same as Fig.~\ref{Fig1sm}, but for the
nonintegrable Ising chain with $h=1$.
}
\label{Fig2sm}
\end{figure}

Figures~\ref{Fig1sm} and \ref{Fig2sm}
compare the differential local variance with the
squared-commutator OTOCs at several probe sites.
These results allow the propagation signatures to be
examined in both integrable and nonintegrable regimes.
The comparison concerns the spatially dependent onset
of the signals rather than their amplitudes, since
the two diagnostics characterize different aspects
of the dynamics.

\begin{thebibliography}{7}
\renewcommand{\baselinestretch}{2.0}

\bibitem{Cheneau2012}
M. Cheneau, P. Barmettler, D. Poletti, M. Endres, P. Schau{\ss},
T. Fukuhara, C. Gross, I. Bloch, C. Kollath, and S. Kuhr,
Light-cone-like spreading of correlations in a quantum many-body system,
Nature {\bf 481}, 484 (2012).

\bibitem{Landsman2019}
K. A. Landsman, C. Figgatt, T. Schuster, N. M. Linke, B. Yoshida,
N. Y. Yao, and C. Monroe,
Verified quantum information scrambling,
Nature {\bf 567}, 61 (2019).

\bibitem{LewisSwan2019}
R. J. Lewis-Swan, A. Safavi-Naini, A. M. Kaufman, and A. M. Rey,
Dynamics of quantum information,
Nat. Rev. Phys. {\bf 1}, 627 (2019).


\bibitem{Polkovnikov2011}
A. Polkovnikov, K. Sengupta, A. Silva, and M. Vengalattore,
Colloquium: Nonequilibrium dynamics of closed interacting
quantum systems,
Rev. Mod. Phys. \textbf{83}, 863 (2011).

\bibitem{Hosur2016}
P. Hosur, X.-L. Qi, D. A. Roberts, and B. Yoshida,
Chaos in quantum channels,
J. High Energy Phys. \textbf{2016}, 004 (2016).


\bibitem{Shenker2014}
S. H. Shenker and D. Stanford,
Black holes and the butterfly effect,
J. High Energy Phys. {\bf 2014}, 67 (2014).

\bibitem{KitaevSuh2018}
A. Kitaev and S. J. Suh,
The soft mode in the Sachdev-Ye-Kitaev model and its gravity dual,
J. High Energy Phys. {\bf 2018}, 183 (2018).

\bibitem{Li2017}
J. Li, R. Fan, H. Wang, B. Ye, B. Zeng, H. Zhai, X. Peng, and J. Du,
Measuring out-of-time-order correlators on a nuclear magnetic resonance
quantum simulator,
Phys. Rev. X {\bf 7}, 031011 (2017).

\bibitem{Garttner2017}
M. G\"arttner, J. G. Bohnet, A. Safavi-Naini, M. L. Wall,
J. J. Bollinger, and A. M. Rey,
Measuring out-of-time-order correlations and multiple quantum spectra
in a trapped-ion quantum magnet,
Nat. Phys. {\bf 13}, 781 (2017).

\bibitem{Swingle2018}
B. Swingle,
Unscrambling the physics of out-of-time-order correlators,
Nat. Phys. {\bf 14}, 988 (2018).

\bibitem{Braum2022}
J. Braum\"uller et al.,
Probing quantum information propagation with out-of-time-ordered correlators,
Nat. Phys. {\bf 18}, 172 (2022).

\bibitem{Qi2026}
X.-L. Qi,
Seeing the quantum butterfly effect,
Physics {\bf 19}, 12 (2026).

\bibitem{LiYC2026}
Y.-C. Li et al.,
Error-resilient reversal of quantum chaotic dynamics enabled by
scramblons,
Phys. Rev. Lett. {\bf 136}, 060403 (2026).

\bibitem{YoshidaYao2019}
B. Yoshida and N. Y. Yao,
Disentangling scrambling and decoherence via quantum teleportation,
Phys. Rev. X {\bf 9}, 011006 (2019).

\bibitem{Harris2022}
J. Harris, B. Yan, and N. A. Sinitsyn,
Benchmarking information scrambling,
Phys. Rev. Lett. {\bf 129}, 050602 (2022).

\bibitem{Vermersch2019}
B. Vermersch, A. Elben, L. M. Sieberer, N. Y. Yao, and P. Zoller,
Probing scrambling using statistical correlations between randomized
measurements,
Phys. Rev. X {\bf 9}, 021061 (2019).

\bibitem{Garcia2021}
R. J. Garcia, Y. Zhou, and A. Jaffe,
Quantum scrambling with classical shadows,
Phys. Rev. Research {\bf 3}, 033155 (2021).

\bibitem{Blocher2024}
P. D. Blocher, K. Chinni, S. Omanakuttan, and P. M. Poggi,
Probing scrambling and operator size distributions using random mixed
states and local measurements,
Phys. Rev. Research {\bf 6}, 013309 (2024).

\bibitem{Wysocki2025}
P. Wysocki and J. Chwede\'nczuk,
Limits to velocity of signal propagation in many-body systems:
A quantum-information perspective,
Phys. Rev. Lett. {\bf 134}, 020201 (2025).

\bibitem{Plodzien2026}
M. P{\l}odzie{\'n} and J. Chwede\'nczuk,
Operator spreading and recoverability of local quantum Fisher
information in a $U(1)$-broken spin chain,
arXiv:2605.02774 (2026).

\bibitem{MartinezAzcona2023}
P. Martinez-Azcona, A. Kundu, A. del Campo, and A. Chenu,
Stochastic operator variance: An observable to diagnose noise and
scrambling,
Phys. Rev. Lett. {\bf 131}, 160202 (2023).

\bibitem{Nahum2018}
A. Nahum, S. Vijay, and J. Haah,
Operator spreading in random unitary circuits,
Phys. Rev. X {\bf 8}, 021014 (2018).

\bibitem{Keyserlingk2018}
C. W. von Keyserlingk, T. Rakovszky, F. Pollmann, and S. L. Sondhi,
Operator hydrodynamics, OTOCs, and entanglement growth in systems
without conservation laws,
Phys. Rev. X {\bf 8}, 021013 (2018).

\bibitem{LiebRobinson1972}
E. H. Lieb and D. W. Robinson,
The finite group velocity of quantum spin systems,
Commun. Math. Phys. {\bf 28}, 251 (1972).

\bibitem{Bravyi2006}
S. Bravyi, M. B. Hastings, and F. Verstraete,
Lieb--Robinson bounds and the generation of correlations and
topological quantum order,
Phys. Rev. Lett. {\bf 97}, 050401 (2006).

\bibitem{Kubo1957}
R. Kubo,
Statistical-mechanical theory of irreversible processes. I.
General theory and simple applications to magnetic and conduction
problems,
J. Phys. Soc. Jpn. {\bf 12}, 570 (1957).

\bibitem{SupplementalMaterial}
See Supplemental Material for the spectral representation
of the response function $\chi(t)$, the derivation of a rigorous bound
relating the differential local variance to a state-dependent squared commutator [Eq.~(\ref{eq:general_variance_otoc})], an exact relation between the local
variance and the XZ-OTOC in the one-excitation sector [Eq.~(\ref{eq:exact_variance_otoc})],
and additional results for integrable and nonintegrable
Ising chains.

\bibitem{Beckman2001}
D. Beckman, D. Gottesman, M. A. Nielsen, and J. Preskill,
Causal and localizable quantum operations,
Phys. Rev. A {\bf 64}, 052309 (2001).

\bibitem{Jurcevic2014}
P. Jurcevic, B. P. Lanyon, P. Hauke, C. Hempel, P. Zoller,
R. Blatt, and C. F. Roos,
Quasiparticle engineering and entanglement propagation in a
quantum many-body system,
Nature {\bf 511}, 202 (2014).

\bibitem{Zhang2017}
J. Zhang et al.,
Observation of a many-body dynamical phase transition with a 53-qubit quantum simulator,
Nature {\bf 551}, 601 (2017).


\bibitem{Luitz2017}
D. J. Luitz and Y. Bar Lev,
Information propagation in isolated quantum systems,
Phys. Rev. B \textbf{96}, 020406(R) (2017).

\bibitem{Khemani2018}
V. Khemani, A. Vishwanath, and D. A. Huse,
Operator spreading and the emergence of dissipative
hydrodynamics under unitary evolution with conservation laws,
Phys. Rev. X \textbf{8}, 031057 (2018).

\bibitem{Auerbach1994}
A. Auerbach,
\textit{Interacting Electrons and Quantum Magnetism}
(Springer, New York, 1994).


\bibitem{Bose2003}
S. Bose,
Quantum communication through an unmodulated spin chain,
Phys. Rev. Lett. {\bf 91}, 207901 (2003).

\bibitem{Fukuhara2013}
T. Fukuhara et al.,
Quantum dynamics of a mobile spin impurity, 
Nat. Phys. {\bf 9}, 235 (2013).

\bibitem{Pfeuty1970}
P. Pfeuty,
The one-dimensional Ising model with a transverse field,
Ann. Phys. (N.Y.) {\bf 57}, 79 (1970).

\bibitem{KimHuse2013}
H. Kim and D. A. Huse,
Ballistic spreading of entanglement in a diffusive
nonintegrable system,
Phys. Rev. Lett. {\bf 111}, 127205 (2013).

\end{thebibliography}

\begin{thebibliography}{7}
\renewcommand{\baselinestretch}{2.0}

\bibitem{Kubo1957}
R. Kubo,
J. Phys. Soc. Jpn. \textbf{12}, 570 (1957).

\bibitem{Auerbach1994}
A. Auerbach,
\textit{Interacting Electrons and Quantum Magnetism}
(Springer, New York, 1994).

\bibitem{LiebRobinson1972}
E. H. Lieb and D. W. Robinson,
Commun. Math. Phys. \textbf{28}, 251 (1972).

\bibitem{CalabreseCardy2006}
P. Calabrese and J. Cardy,
Phys. Rev. Lett. \textbf{96}, 136801 (2006).

\bibitem{Khemani2018}
V. Khemani, A. Vishwanath, and D. A. Huse,
Phys. Rev. X \textbf{8}, 031057 (2018).

\end{thebibliography}
\end{document}